\documentclass{article}
\usepackage{spconf,amsmath,graphicx}

\usepackage{enumitem}
\setlist{nosep, leftmargin=14pt}

\usepackage{mwe} 
\usepackage{amsthm}
\usepackage{algorithmicx,algorithm}
\usepackage{amsbsy}
\usepackage{amsfonts}
\usepackage{marvosym}
\usepackage{amsmath}
\usepackage{threeparttable}
\usepackage{booktabs}
\usepackage{multirow}
\usepackage{float}
\usepackage{url}

\title{A resource-efficient deep learning framework for low-dose brain PET image reconstruction and analysis}
\name{\begin{tabular}{c}Yu Fu$^{1,2}$\sthanks{These authors contributed equally to this work.}, Shunjie Dong$^{1\ast}$, Yi Liao$^{1}$, Le Xue$^{1}$, Yuanfan Xu$^{3}$, Feng Li$^{3}$, \\ Qianqian Yang$^{1}$, Tianbai Yu$^{1}$, Mei Tian$^{1}$\sthanks{Corresponding authors.} and Cheng Zhuo$^{1\dagger}$\end{tabular}}

\address{$^{1}$ Zhejiang University, Hangzhou, China\\
$^{2}$ Binjiang Institute of Zhejiang University, Hangzhou, China\\
$^{3}$ Hangzhou Universal Medical Imaging Diagnostic Center, Hangzhou, China}

\begin{document}
%
\maketitle
\begin{abstract}
$^{18}$F-fluorodeoxyglucose (18F-FDG) Positron Emission Tomography (PET) imaging usually needs a full-dose radioactive tracer to obtain satisfactory diagnostic results, which raises concerns about the potential health risks of radiation exposure, especially for pediatric patients. Reconstructing the low-dose PET (L-PET) images to the high-quality full-dose PET (F-PET) ones is an effective way that both reduces the radiation exposure and remains diagnostic accuracy. In this paper, we propose a resource-efficient deep learning framework for L-PET reconstruction and analysis, referred to as \textit{transGAN-SDAM}, to generate F-PET from corresponding L-PET, and quantify the standard uptake value ratios (SUVRs) of these generated F-PET at whole brain. The transGAN-SDAM consists of two modules: a transformer-encoded Generative Adversarial Network (transGAN) and a Spatial Deformable Aggregation Module (SDAM). The transGAN generates higher quality F-PET images, and then the SDAM integrates the spatial information of a sequence of generated F-PET slices to synthesize whole-brain F-PET images. Experimental results demonstrate the superiority and rationality of our approach.
\end{abstract}
\begin{keywords}
Image reconstruction, Deep learning, Positron Emission Tomography
\end{keywords}

\vspace{-0.5cm}

\section{Introduction}
\label{sec:intro}

Positron Emission Tomography (PET) is the main neuroimaging method in the diagnosis of pediatric epilepsy (PE)~\cite{von2018pet,juhasz2020utility}. A full-dose radioactive tracer may raise concerns about the potential health risks for PE subjects. Hence, researchers have tried to reduce the tracer dose as much as possible for the PET scans to mitigate this negative effect~\cite{tan2020total}. However, the quality of low-dose PET (L-PET) images, containing more noises and artifacts, is inevitably lower than that of full-dose PET (F-PET) images, which hardly meets the requirements for satisfactory diagnosis~\cite{alotaibi2020diagnostic}.

A potential solution to solve the above problem is to reconstruct full-dose-like PET images from the L-PET ones. Some recent studies~\cite{kaplan2019full,chen2019ultra,ouyang2019ultra,lei2019whole,spuhler2020full} proposed deep learning methods to reconstruct L-PET images, achieving high visual quality for the physicians to conduct a diagnosis. However, to our knowledge, there are still two critical issues that have not been well addressed: 
(1) How to further strengthen the understanding of the semantic patterns between voxels during L-PET reconstruction? (2) How to trade off the accuracy, speed, and resource consumption during L-PET reconstruction?

To address these issues, we propose a resource-efficient framework for L-PET reconstruction, referred to as \textit{transGAN-SDAM}. The \textit{transGAN-SDAM} generates F-PET images from L-PET images in a fast way, and utilizes the spatial information of generated F-PET image slices to refine the final 3D F-PET images for a more accurate standard uptake value ratio (SUVR) analysis of the whole brain. The transGAN-SDAM consists of two modules: a transformer-encoded Generative Adversarial Network (transGAN) and a Spatial Deformable Aggregation Module (SDAM). The transGAN optimizes the reconstruction process from L-PET to F-PET and generates high-quality F-PET image slices by introducing a transformer-encoded generator. The SDAM takes a sequence of generated F-PET slices as the inputs and extracts their spatial information by an offset prediction network. The spatial information is then further refined via a deformable convolution to optimize all generated F-PET images in a slice-by-slice manner. Finally, these refined F-PET image slices are combined to form the final 3D PET images for quantitative analysis.
\vspace{-0.5cm}

\begin{figure*}[t]
 \centering
 \centerline{\includegraphics[width=0.9\textwidth]{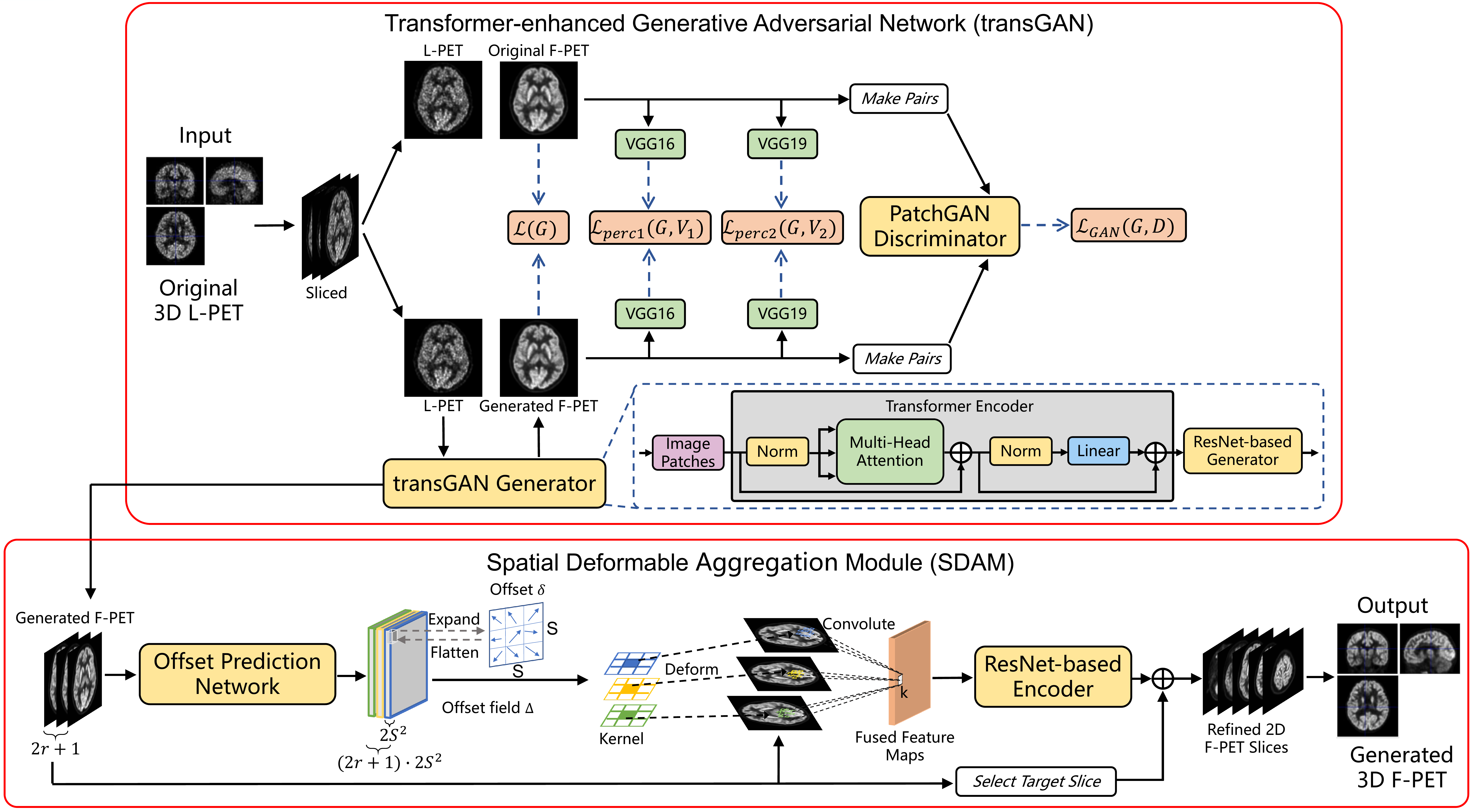}}
\caption{The overall architecture of transGAN-SDAM. Our learning procedure consists of two steps: we first train transGAN that generates F-PET images and then update SDAM to refine the generated F-PET images for SUVR analysis of the whole brain.} 
\label{whole_architecture}
\end{figure*}

\section{Datasets}
\label{sec:data}
The experimental dataset consists of 45 pediatric subjects with epilepsy, which are scanned and collected in 2020 at Hangzhou Universal Medical Imaging Diagnostic Center~\footnote{http://www.uvclinic.cn/universal-hangzhou/}. All subjects’ brain FDG-PET images are acquired on a whole-body hybrid PET/MR system (SIGNA PET/MR, GE Healthcare). Considering the clinical practice, we did not exclude those images with relatively lower quality.
\vspace{-0.5cm}

\section{Methology}
\label{sec:metho}

\subsection{Image preprocessing}
L-PET images are reconstructed through 5$\%$ undersampling of list-mode F-PET data. Both L-PET images and F-PET images are preprocessed by Statistical Parametric Mapping (\url{https://www.fil.ion.ucl.ac.uk/spm/}) for realignment and normalization. After preprocessing, the voxel size of these PET images is $1\times 1\times 1$ mm, and each 3D PET image is split into several closely stacked 2D PET image slices with $256\times256$ pixels. 

\begin{figure*}[t]
 \centering
 \centerline{\includegraphics[width=16cm]{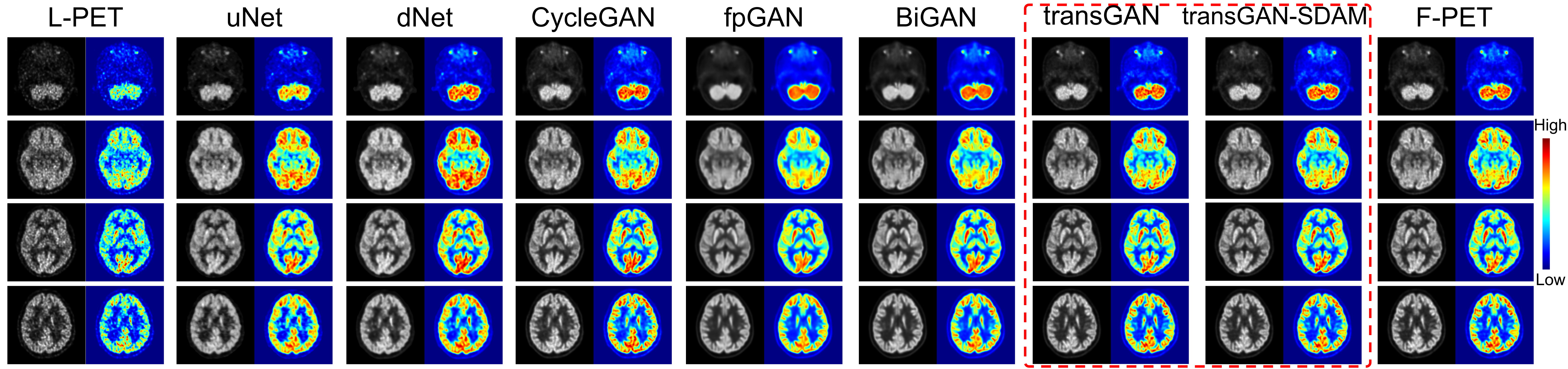}}
\caption{Visualization of the L-PET images, the synthesized F-PET images by different methods and the original F-PET images (gray-scale and colorful). Results of transGAN and transGAN-SDAM are denoted by the red dashed line. The colors from blue to red represent metabolic values from low to high.} 
\label{results_comparison}
\end{figure*}

\subsection{transGAN}
The proposed transGAN-SDAM consists of transGAN and SDAM, as illustrated in Fig.~\ref{whole_architecture}. Firstly, L-PET image slices are fed into transGAN, containing four sub-networks: transGAN Generator $G$, PatchGAN Discriminator $D$~\cite{kaplan2019full}, high feature encoder VGG16-Net and VGG19-Net~\cite{simonyan2014very}, to generate the corresponding F-PET slices in a 2.5D manner. Different from previous GAN-based studies that directly adopting a uNet generator~\cite{isola2017image} or a ResNet generator~\cite{dar2019image,hu2020brain} for image generation, we design a transformer-encoded ResNet generator $G$ that can encode and optimize all image patches in whole-slice scale through the multi-headed self-attention mechanism before every generation process. Since all generation processes are pre-encoded, the transGAN converges more quickly and ensure the semantic stability of generated F-PET images. During training the proposed transGAN, the adversarial loss function $\mathcal{L}_{GAN}$ is defined as:
\begin{small}
\begin{equation}
  \mathcal{L}_{GAN}(D,G) =\\ -\mathbb{E}_{x,y}[(D(x,y)-1)^2]-\mathbb{E}_{x}[D(x,G(x))^2],
\end{equation}
\end{small}
where $x$ denotes the L-PET slices, $G(x)$ denotes the F-PET image generated by the generator, $y$ is the corresponding ground-truth F-PET image and $\mathbb{E}$ represents the expectation. We also introduce the Charbonnier Loss~\cite{barron2019general} to punish the Euclidean difference between the synthesized F-PET images and the ground truth: 
\begin{small}
\begin{equation}
\label{gan}
  \mathcal{L}(G) = \mathbb{E}_{x,y}[\sqrt{\Arrowvert y-G(x) \Arrowvert^2+\epsilon^2}],
\end{equation}
\end{small}
where $\epsilon$ denotes a small constant that is set to be $1\times10^{-8}$. Considering the perceptual difference between the generated and the ground-truth F-PET images, inspired by~\cite{ouyang2019ultra,dar2019image}, we include both a VGG16-Net and a VGG19-Net that are pretrained on the ImageNet~\cite{deng2009imagenet} to extract high feature representations of both $y$ and $G(x)$. The dual perceptual loss is defined as:
\begin{small}
\begin{equation}
\label{perc}
\begin{array}{r}
  \mathcal{L}_{perc}(G,V) = \mathbb{E}_{x,y}[\sqrt{\Arrowvert V_{1}(y)-V_{1}(G(x)) \Arrowvert^2+\epsilon^2}]\\
  +\mathbb{E}_{x,y}[\sqrt{\Arrowvert V_{2}(y)-V_{2}(G(x)) \Arrowvert^2+\epsilon^2}],
\end{array}
\end{equation}
\end{small}
where $V_{1}(\cdot)$ and $V_{2}(\cdot)$ represent the feature maps extracted by VGG16-Net and VGG19-Net, respectively. Thus, the total loss of training transGAN is given as follows:
\begin{small}
\begin{equation}
\label{loss_transGAN}
\mathcal{L}_{\text {trans}}(G,V,D)=\mathcal{L}_{GAN}(D,G)+\alpha \mathcal{L}(G)+\beta \mathcal{L}_{\text {perc}}\left(G,V\right),
\end{equation}
\end{small}
where $\alpha$ and $\beta$ are the hyper-parameters that govern the trade off among $\mathcal{L}_{GAN}(D,G)$, $\mathcal{L}(G)$ and $\mathcal{L}_{perc}(G,V)$. The hyper-parameters $\alpha$ and $\beta$ are set to be 100 and 100, respectively.

\begin{figure*}[tb]
 \centering
 \centerline{\includegraphics[width=16cm]{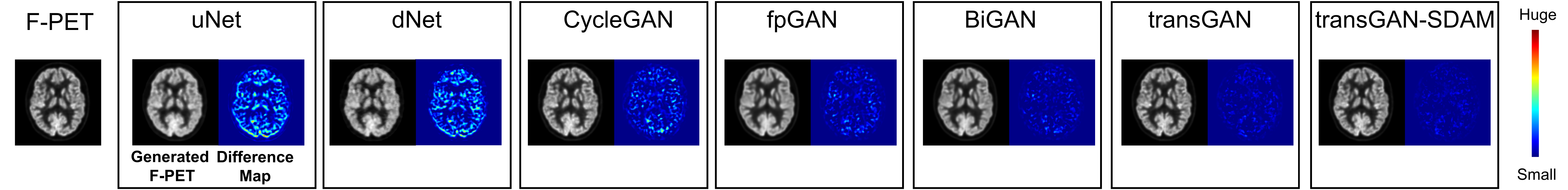}}
\caption{Comparison of the pseudo-color difference maps between the real and the generated F-PET images by compared methods and our proposal. The absolute differences from small to huge are represented by the color from blue to red.} 
\label{difference_maps}
\end{figure*}

\subsection{SDAM}
To generate 3D F-PET images, we propose SDAM to better extract additional informative structural contents and semantic details with a deformable convolution layer~\cite{dai2017deformable,dong2020deu}. Different from existing methods~\cite{tian2020tdan,dong2021rconet} that explored spatial information by neighboring slices in a pairwise manner, we employ a U-Net based network as the offset prediction network which is fed into a sequence of generated F-PET images, to predict all the deformable offset field $\Delta$~\textit{jointly} for the target slice that needs be enhanced, as shown in Fig.~\ref{whole_architecture}. It demonstrates that the joint manner has a much lower computational cost than the pairwise scheme~\cite{tian2020tdan} since all the deformation offsets can be obtained in a single forward pass. 
Then, the offset fields are further adaptively fused into the sequence of generated F-PET images via a deformable convolutional layer to get fused feature maps:
\begin{small}
\begin{equation}
\label{SDAM}
  \mathcal{F}(\mathbf{p}) = \sum_{t={t_{0}-r}}^{t_{0}+r}\sum_{s=1}^{S^2}K_{t,s}\cdot C_t(\mathbf{p}+\mathbf{p}_s+\mathbf{\delta}_{(t,\mathbf{p}),s}),
\end{equation}
\end{small}
where $\mathcal{F}(\cdot)$ denotes the derived feature map, $S^2$ represents the size of convolution kernel and $K_t \in \mathbb{R}^{S^2}$ denotes the kernel for $t$-th channel. $\mathbf{p}$ is a spatial position and $\mathbf{p}_s$ indicates the regular sampling offset. the deformation offsets $\mathbf{\delta}_{(t,\mathbf{p}),s}$ are position-specific, i.e., an individual $\mathbf{\delta}_{(t,\mathbf{p}),s}$ will be assigned to each convolution window centered at the spatial position $(t, \textbf{p})$.
In particular, the preceding and succeeding $r$ slices are regarded as the reference, which together with the target slice $C_{t_{0}}$ input to SDAM. Thus, the offset field $\Delta \in \mathbb{R}^{(2r+1) \times 2S^2 \times H \times W}$ for all the spatial positions in the sequence of F-PET slices:
\begin{small}
\begin{equation}
\Delta =  \mathcal{F}_{\theta_{os}}([C_{t_{0}-r},\cdots,C_{t_{0}},\cdots,C_{t_{0}+r}]),
\end{equation}
\end{small}
where $[\cdot]$ denotes the concatenation of the generated F-PET slices, and $\mathcal{F}_{\theta_{os}}(\cdot)$ is the offset prediction network.
$H$, $W$ is the height and width, respectively. 

To fully explore the complementary information in the fused feature maps, we feed $\mathcal{F}$ through a ResNet-based reconstruction network to predict a restoration residue $\hat{C}_{t_0}^{HQ}$. The refined target slice $C_{t_0}^{HQ}$ can finally be obtained by adding the residue back to the target slice as:
\begin{small}
\begin{equation}
C_{t_0}^{HQ} = \hat{C}_{t_0}^{HQ} + C_{t_0}.
\end{equation}
\end{small}
Thus, the loss function $\mathcal{L}_{SDAM}$ is set to the sum of squared error between the refined target frame $C_{t_0}^{HQ}$ and the ground-truth $y$: 
\begin{small}
\begin{equation}
\label{loss}
\mathcal{L}_{SDAM} = {\Arrowvert y - C_{t_0}^{HQ} \Arrowvert}^2_2.
\end{equation}
\end{small}
Finally, each subject's all refined F-PET image slices are combined together to form the final 3D PET images for quantitative evaluations such as SUVR analysis.

\begin{table*}[t]\small
 \centering
 \caption{Quantitative comparisons of transGAN and other approaches for L-PET reconstruction (mean $\pm$ std).}
 \label{table1}
 \begin{threeparttable}
 \begin{tabular}{lccccccc}  
  \toprule   
  Method & uNet~\cite{chen2019mmdetection} & dNet~\cite{spuhler2020full} & CycleGAN~\cite{lei2019whole} & fpGAN~\cite{ouyang2019ultra} & BiGAN~\cite{hu2020brain} & transGAN & transGAN-SDAM \\
  \midrule   
  PSNR & 26.1$\pm$1.6 & 25.8$\pm$1.6 & 28.4$\pm$1.8 & 30.7$\pm$2.2 & 31.9$\pm$3.2 & 32.9$\pm$4.3 & {\bf 33.9$\pm$4.0}  \\  
  SSIM & 0.861$\pm$0.02 & 0.876$\pm$0.02 & 0.898$\pm$0.02 & 0.929$\pm$0.02 & 0.939$\pm$0.03 & 0.946$\pm$0.04 & {\bf 0.950$\pm$0.04}  \\  
  VSMD & 0.133$\pm$0.077 & 0.148$\pm$0.069 & 0.099$\pm$0.065 & 0.057$\pm$0.036 & 0.057$\pm$0.040 & 0.045$\pm$0.037& {\bf 0.043$\pm$0.039}  \\
  \bottomrule  
 \end{tabular}
 \end{threeparttable}
\end{table*}

\section{Experiments and Results}
\label{sec:results}
The average peak signal-to-noise ratio (PSNR), structural similarity index measurement (SSIM), and voxel-scale metabolic difference (VSMD) are adopted in this paper to quantitatively evaluate the generated F-PET slices and original F-PET slices. Besides, the SUVRs are calculated to describe the numerical stability of refined F-PET images by our framework relative to original 3D F-PET images. We compared the proposed transGAN-SDAM with 6 state-of-the-art models that are specially designed for L-PET reconstruction or medical image synthesis: uNet~\cite{chen2019mmdetection}, dNet~\cite{spuhler2020full}, CycleGAN~\cite{lei2019whole}, fpGAN~\cite{ouyang2019ultra}, BiGAN~\cite{hu2020brain} and 3D CycleGAN~\cite{zhao2020study}. To avoid overfitting, we conducted the 10-fold cross-validation, where 8 folds are the training set, 1 fold is the validation set and remaining 1 fold as the test set. Due to the space limitation, the more implementation details, experimental setting, and some results of the comparison are shown in our supplementary materials.

\subsection{Quantitative analysis}

As shown in Table~\ref{table1}, we can observe that transGAN-SDAM significantly outperforms all of the competing models (\textit{p}-values $<$ 0.05 for all paired comparisons) with the highest PSNR (33.9$\pm$4.0), SSIM (0.950$\pm$0.04), and the lowest VSMD (0.043$\pm$0.039), indicating that the quality of generated F-PET images by transGAN-SDAM is closest to the ground-truth.
It is worth noting that the transGAN-SDAM module substantially improves PSNR by at least 2.0 dB, SSIM by at least $1.1\%$, and decreases the VSMD by at least $24.6\%$ over these state-of-the-art approaches. We also find that SDAM can further refine the image quality to achieve higher PSNR, SSIM and lower VSMD when comparing transGAN and transGAN-SDAM. As shown in Fig.~\ref{results_comparison}, we observe that transGAN and transGAN-SDAM can produce more accurate structural and metabolic details than compared approaches. The pseudo-color difference maps between the real and the generated F-PET images by different approaches are illustrated in Fig.~\ref{difference_maps}, which shows that transGAN-SDAM achieves the smallest voxel-scale difference to the ground-truth F-PET images.

\begin{figure}[t]
\centering
\centerline{\includegraphics[width=8cm]{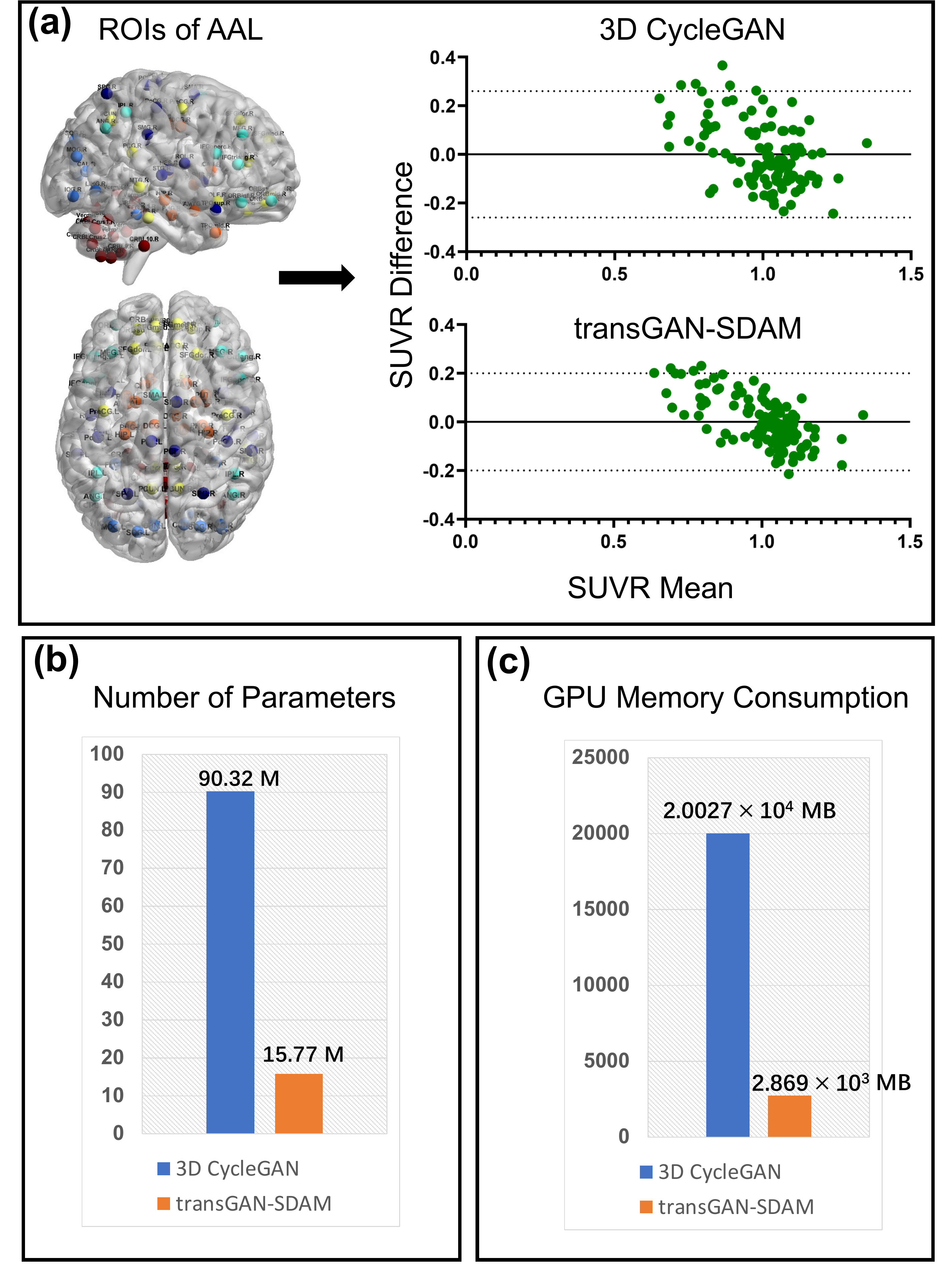}}
\caption{(a) Comparisons of Bland-Altman two-dimensional point diagrams of SUVRs for the same subject. Each green point denotes an SUVR to its corresponding brain region (ROI) in AAL brain template. (b) Number of parameters. (c) GPU memory consumption.}
\label{trans_vs_CycleGAN3D}
\end{figure}

\subsection{Efficiency and SUVR analysis}
We further present the performance comparison between transGAN-SDAM and 3D CycleGAN~\cite{oulbacha2020mri} in Fig.~\ref{trans_vs_CycleGAN3D}. As show in Fig.~\ref{trans_vs_CycleGAN3D} (a), we can observe that whole-brain SUVRs of transGAN-SDAM have smaller 95$\%$ limits of agreements (from -0.19 to 0.19), smaller 95$\%$ CI (from -0.018 to 0.018) and stronger correlation coefficient (0.831) than 3D CycleGAN (agreement from -0.26 to 0.26, CI from -0.024 to 0.024, correlation coefficient: 0.633). Meanwhile, 3D CycleGAN needs 90.32 M of parameters to train (Fig.~\ref{trans_vs_CycleGAN3D} (b)) and $2.0027\times10^{4}$ MB of GPU memory to implement (Fig.~\ref{trans_vs_CycleGAN3D} (c)), which means much higher resource consumption than our transGAN-SDAM (15.77 M of parameters to train and $2.869\times10^{3}$ MB of GPU memory to implement). It indicates that compared with 3D-based models, our proposal can satisfy specific needs, e.g., high response speed and limited GPU resources.

\section{Conclusion}
\label{sec:discuss}

In this paper, we propose the \textit{transGAN-SDAM} as a resource-efficient framework for fast 2.5D-based L-PET reconstruction and accurate whole-brain quantitative analysis. 
Experimental results show that our proposal achieves the state-of-the-art performance on commonly used metrics, indicating the effectiveness and rationality of combining both transGAN and SDAM.
This framework could facilitate the wider adoption of deep learning methods into clinical workflows.

\section{Compliance with Ethical Standards}
\label{sec:ethical}

The study design and exemption from informed consent were approved by the Institutional Review Board of Hangzhou Universal Medical Imaging Diagnostic Center (Approval No. [2021] 001).

\section{Acknowledgements}
\label{sec:acknowledgments}

This work was supported in part by the grant from the National Science Foundation of China (No. 62034007), in part by the Zhejiang Provincial Innovation Team Project (No. 2020R01001), and in part by the Zhejiang Lab’s International Talent Fund for Young Professionals (ZJ2020JS013).




\bibliographystyle{IEEEbib}
\fontsize{9pt}{10pt} 
\selectfont
\bibliography{strings,refs}

\end{document}